\documentclass[twoside,12pt]{article}

\usepackage[dvips]{graphics,epsfig}
\usepackage[latin1]{inputenc}
\usepackage{setspace}
\usepackage{amsfonts}

\usepackage{textcomp}
\usepackage{amsmath}
\usepackage{graphicx}
\usepackage{epstopdf}

\usepackage{subfig}

\def\laq{\raise 0.4 ex \hbox{$<$}\kern -0.8 em\lower 0.62 ex\hbox{$\sim$}}
\def\gaq{\raise 0.4 ex \hbox{$>$}\kern -0.7 em\lower 0.62 ex\hbox{$\sim$}}

\def\JP{{\it J. Phys.} }

\def\PR{{\it Phys. Rev.} }

\def\RMP{{\it Rev. Mod. Phys.} }

\def\beq{\begin{equation}}
\def\eeq{\end{equation}}
\def\beqa{\begin{eqnarray}} 
\def\eeqa{\end{eqnarray}}

\begin{document}
\pagestyle{plain}

\begin{flushright}
{\bf ACCEPTED VERSION }\\
April 24, 2020
\end{flushright}
\vspace{15mm}

\begin{center}

{\Large\bf Functional equations for regularized zeta-functions and diffusion processes}

\vspace*{1.0cm}

Alexis Saldivar$^{*}$, Nami F. Svaiter$^{**}$\\
\vspace*{0.5cm}
{Centro Brasileiro de Pesquisas F\'{\i}sicas\\
Rua Xavier Sigaud, 150 - Urca, Rio de Janeiro - RJ, 22290-180, Brazil}\\

\vspace*{1.0cm}

Carlos A. D. Zarro$^{***}$\\
\vspace*{0.5cm}
{Instituto de F\'isica\\
 Universidade Federal do Rio de Janeiro,\\
Av. Athos da Silveira Ramos, 149 - Cidade Universit\'aria da Universidade Federal do Rio de Janeiro, Rio de Janeiro - RJ, 21941-909, Brazil}\\

\vspace*{2.0cm}
\end{center}

\begin{abstract}
We discuss modifications in the integral representation of the Riemann zeta-function that lead to generalizations of the Riemann functional equation that preserves the symmetry $s\to (1-s)$  in the critical strip. By modifying one integral representation of the zeta-function with a cut-off that does exhibit the symmetry $x\mapsto 1/x$, we obtain a generalized functional equation involving Bessel functions of second kind. Next, with another cut-off that does exhibit the same symmetry, we obtain a generalization for the functional equation involving only one Bessel function of second kind. Some connection between one regularized zeta-function and the Laplace transform of the heat kernel for the Euclidean and hyperbolic space is discussed.
\end{abstract}

\vspace{2cm}
\begin{flushright}
\textit{Accepted for publication in Journal of Physics A.}
 \end{flushright}

\vfill
\noindent\underline{\hskip 140pt}\\[4pt]
{$^{*}$ E-mail address: asaldivar@cbpf.br} \\
\noindent
{$^{**}$ E-mail address: nfuxsvai@cbpf.br} \\
\noindent
{$^{***}$ E-mail address: carlos.zarro@if.ufrj.br}

\newpage


\onehalfspacing

\section{Introduction}
\quad
The Riemann approach with a zeta-function uses   
complex variable tools 
to discuss the distribution of primes \cite{riem}. Riemann obtained two results: the analytic extension of the zeta-function and that this zeta-function satisfies a functional equation. He presented two proofs for this functional equation. The first one using the theta function and its Melin transform and the second one using countour integration. 
Riemann also made a still unproved conjecture  that the non-trivial complex zeros of the zeta-function must all lie on the critical line $\Re\,(s)=\frac{1}{2}$ \cite{hardy,titchmarsh,ingham}. 
On way to discuss the Riemann conjecture is to study different functions in number theory, where it is possible to establish an analog of the Riemann hypothesis. This is known in the literature as the generalized Riemann hypothesis. In this scenario there are several ways to proceed, a possible one is to investigate Dirichlet L-functions \cite{livro}. These functions have analytic continuation to the whole complex plane and also satisfy functional equations. A quite distinct approach is to approximate the Riemann zeta-function by some simpler function and investigate the dynamics of the zeros of these functions when some parameter changes \cite{hejhal}. There is also an approach discussing approximate representation of the Riemann zeta-function in terms of finite sums \cite{hl}. This approach leads to an approximate functional equation that possesses an interesting symmetry in the critical line. This symmetry allows one to investigate the distribution of the non-trivial zeros. With this case in hand, the distribution of the zeros of partial sums of the Riemann zeta-function  has been investigated. See, $e.g.$, references \cite{alexa,gonek}.

The aim of this paper is twofold. First, is to discuss modified zeta-functions and prove that the these zeta-functions satisfy generalized Riemann functional equations. 
Second, we discuss the connection between one modified zeta-function and the Laplace transform of the heat kernel for the Euclidean space $\mathbb{R}^{d}$ and a $d$-dimensional hyperbolic space $\mathbb{H}^{d}$. In order to obtain the analytic extension of the zeta-function, Riemann discussed a special case of the Jacobi theta-function, a function with modular symmetry that satisfies a functional equation. As it has been stressed in the literature, all reasonable generalizations of the Riemann zeta-function must be related to a modular form \cite{brown}. Here, we discuss the consequences  of introducing, in the integral representation of the zeta-function, cut-offs  that are invariant under the  transformation $x\mapsto 1/x$. We prove that
these modified zeta-functions lead to generalizations for the Riemann functional equation with the same symmetry $s\to (1-s)$  in the critical strip. Next,  we modify one integral representation of the zeta-function to obtain functional equations where modified Bessel functions of second kind appear. 
We also discuss modified zeta-function with a cut-off defined by two parameters. We show a connection between Laplace transform of the heat kernel for the Euclidean space $\mathbb{R}^{d}$ and $\mathbb{H}^{d}$
and this modified zeta-function.
The importance of the transformation $x\mapsto 1/x$ to investigate connections between number theory and physics was pointed out in reference \cite{sierra}. In references \cite{lagariassuzuki,ki}, the Riemann hypothesis for certain integrals of Eisenstein series was discussed.

Before concluding, we would like to point out some connections between number theory and quantum field theory, known as arithmetic quantum theory \cite{stn,spec1,bakas,spec2,spec3}, that have been discussed in the literature. A classical example of such system is a bosonic Riemann gas, a second quantized mechanical system at temperature $\beta^{-1}$ with partition function given by the Riemann zeta-function. Recently, it was discussed  connections between the non-trivial zeros of the Riemann zeta-function and the behavior of the thermodynamic free energy of a disordered system using the arithmetic bosonic gas with quenched disorder \cite{prime5}. Introducing randomness in the bosonic Riemann gas and studying the thermodynamic variables of this arithmetic gas in the complex $\beta$-plane, the connection between the zeros of the Riemann zeta-function and physics was established.  For other papers discussing formal aspects of quantum field theory and number theory see the references \cite{mussardo,connes,rosu,zyl,pre,prime1,prime2,Frontiers,Daniel,prime3,prime4}.

The organization of the paper is the following. In section \ref{Sec:Riem}, we briefly discuss some properties of the Riemann zeta-function. In section \ref{Sec:modified} we discuss modified zeta-functions. In section \ref{Sec:modified2} we present a modified zeta-function with a cut-off that does exhibit the symmetry $x\mapsto 1/x$.  We obtain a generalization for the Riemann functional equation with the same symmetry $s\to (1-s)$ in the critical strip, involving two Bessel functions of second kind. In section \ref{Sec:modified4}, we obtain a generalization for the functional equation involving only one modified Bessel function of second kind. In section \ref{Sec:modified3} the similarities  between one of the modified zeta-function and the mathematics of diffusion processes are discussed. In section \ref{Sec:modified33} some connection between a regularized zeta-function and the Laplace transform of the heat kernel for Euclidean and hyperbolic space are discussed. Conclusions are given in section \ref{Sec:Conclusion and perspectives}.   

\section{The Riemann zeta-function} \label{Sec:Riem}

The Riemann zeta-function $\zeta(s)$ is a function of the complex variable $s=\sigma+i t$, where $\sigma, t\,\, \in{\mathbb{R}}$. It is defined in the half-plane $\Re\,(s)>1$
by the Euler product, where the product is taken over all prime numbers $p\in {\mathbb{P}}$,
or by the set of natural numbers $n\in {\mathbb{N}}$, by a Dirichlet series. We have
\begin{equation}
\zeta(s)=\prod_{p}\,\Bigl(1-{p^{-s}}\Bigr)^{-1}=\sum_{n=1}^{\infty}\,\frac{1}{n^{s}}.
\label{p2}
\end{equation}
The series defined above is absolutely convergent and therefore the sum is analytic in the half-plane $\Re\,(s)>1$, and 
can be extended to the complex plane 
as a meromorphic function with a simple pole at $s=1$.  Here we will briefly discuss one way to perform the analytic extension of the Riemann zeta-function. Starting from the definition of the gamma function and defining the function $\psi(x)$ as
\begin{equation}
\psi(x)= \sum_{n=1}^{\infty}\,e^{-n^{2}\,\pi\,x},
\label{z2}
\end{equation}
we can write 
\begin{equation}
\pi^{-\frac{s}{2}}\,\Gamma\biggl(\frac{s}{2}\biggr)\zeta(s)=
\int_{0}^{\infty}dx\,\psi(x)\,x^{\frac{s}{2}-1}. \label{11}
\end{equation}
Following Riemann, we split the integral representation defined in equation (\ref{11}) into two integrals. We get
\begin{equation}\label{z3}
\pi^{-\frac{s}{2}}\,\Gamma\biggl(\frac{s}{2}\biggr)\zeta(s)=
\int_{0}^{1}dx\,\psi(x)\,x^{\frac{1}{2}(s-2)}+\int_{1}^{\infty}dx\,\psi(x)\,x^{\frac{1}{2}(s-2)}. 
\end{equation}
In order to proceed, one can use the Poisson summation formula. This is an important point, since one way to obtain the analytic extension of the Riemann zeta-function from $\Re\,(s)>1$
to the whole complex plane, is using the properties of the function $\Theta:\, (0,\infty)\rightarrow (0,\infty)$ defined by the formula
\begin{equation}\label{theta}
\Theta(v)=\sum_{n \in\,\mathbb{Z}}\,e^{-\pi\,n^{2}\,v}. 
\end{equation}
This $\Theta(v)$ is a particular case of the Jacobi theta-function i.e., $\theta(t)=\vartheta_{3}(0,e^{-\pi t})$ where the Jacobi theta-function is defined as
\begin{equation}
\vartheta_{3}(z,q)=\sum_{n \in\,\mathbb{Z}}\,e^{2\pi i nz}q^{n^{2}}=1+\sum_{n=1}^{\infty}q^{n^{2}}\,\cos 2\pi nz.
\end{equation}
Using Poisson summation formula one can show that the function $\Theta(v)$ satisfies the functional equation 
$\Theta(\frac{1}{v})=\sqrt{v}\,\Theta(v)$. A consequence for this identity is the functional equation for the Riemann zeta-function. 
Since this function is related to the  function $\psi(x)$ defined in equation (\ref{z2}) by $\psi(v)=\frac{1}{2}(\Theta(v)-1)$,  using the identity $\psi(x)=-\frac{1}{2}+\frac{1}{\sqrt x}(\psi(\frac{1}{x})+\frac{1}{2})$  one obtain
\begin{equation}
\pi^{-\frac{s}{2}}\,\Gamma\biggl(\frac{s}{2}\biggr)\zeta(s)=
\frac{1}{s(s-1)}+\int_{1}^{\infty}dx\,\psi(x)\biggl
(x^{\frac{1}{2}(s-2)}+x^{-\frac{1}{2}(s+1)}\Biggr).
\label{z7}
\end{equation}
The integral that appears in  equation (\ref{z7}) is convergent for all values of $s$ and therefore  equation (\ref{z7}) gives the analytic continuation of the Riemann zeta-function to the whole complex $s$-plane. The only singularity is the pole at $s=1$. It is possible to show that Riemann zeta-function $\zeta(s)$ satisfies the functional equation
\begin{equation}\label{eq:Riemannoriginal}
\pi^{-\frac{s}{2}}\Gamma\biggl(\frac{s}{2}\biggr)\zeta(s)=\pi^{-\frac{(1-s)}{2}}
\Gamma\biggl(\frac{1-s}{2}\biggr)\zeta(1-s),
\end{equation}
valid for $s\in {\mathbb{C}} \setminus\left\{0,1\right\}$. This is a quite important equation, since it connects two functions outside the original domain of convergence. Other equivalent way to write the functional equation using the properties of the gamma function is to define
\begin{equation}\label{ef}
\vartheta(s)=\frac{(2\pi)^{s}}{2\Gamma(s)\cos(\frac{\pi\,s}{2})}.
\end{equation}
Therefore we have $\zeta(s)=\vartheta(s)\zeta(1-s)$. Defining the entire function $\xi(s)$ as 
\begin{equation}
\xi(s)=s(s-1)\pi^{-\frac{s}{2}}\Gamma\biggl(\frac{s}{2}\biggr)\zeta(s),
\end{equation}
the functional equation becomes $\xi(s)=\xi(1-s)$. In the next section we discuss modifications in the integral representation of the Riemann zeta-function introducing functions that exhibit the symmetry $x\mapsto 1/x$. This procedure lead us to generalizations for the Riemann functional equation with the same symmetry $s\to (1-s)$ in the critical strip.

\section{Modified zeta-functions} \label{Sec:modified}

As we mention, one way to make progress in the Riemann problem is to consider families of Dirichlet L-functions. There is a quite important function concerning the study of prime numbers in arithmetic progressions with difference $m$. Let $m$ a natural number and let $\chi(m)$ be a Dirichlet character modulo $m$. The function $L(s,\chi)$ for $s=\sigma+it$, $\sigma>1$ is defined as 
\begin{equation}
L(s,\chi)=\sum_{n=1}^\infty\,\frac{\chi(m)}{n^{s}}.
\end{equation}
This function is holomorphic for $\Re\,(s)>1$.
For $\sigma>1$ there is an Euler product identity
\begin{equation}
L(s,\chi)=\prod_{p}\bigl(1-\chi(p)p^{-s}\bigr)^{-1},
\end{equation}
that can be proved using a multiplicativity of the $\chi$ function.
In the following, we would like to discuss different kinds of modified zeta-functions. One well known modification in the series representation of the zeta-function is the introduction of a sharp cut-off. 

One should start from a classical result \cite{hl}. Using an approximate representation of the zeta-function in terms of finite sums, it was proved that
\begin{equation}
\zeta(s)=\sum_{n\leq\,x} \frac{1}{n^{s}}+\vartheta(s)\sum_{n\leq\,y}\frac{1}{n^{1-s}}+O(x^{-\sigma})+O(t^{\frac{1}{2}-\sigma}\,y^{\sigma-1}),
\label{zz444}
\end{equation}
for $0\leq\,\sigma<1$ and $x,y,t >C>0$ and $xy=\frac{t}{2\pi}$. This is known as an approximate functional equation. This function $\vartheta(s)$ presented in equation (\ref{ef}) can be used to define 
the Hardy function, to investigate the distribution of non-trivial zeros in the critical line \cite{nn}. This function is defined as $Z(t)=\zeta(\frac{1}{2}+it)\bigl(\vartheta(\frac{1}{2}+it)\bigr)^{-\frac{1}{2}}$. 
Using that $(\vartheta(\frac{1}{2}+it))^{*}=
\vartheta(\frac{1}{2}-it)=\vartheta^{-1}(\frac{1}{2}+it)$,  so that $Z(t)\in {\mathbb{R}}$ when $t \in {\mathbb{R}}$, we get that $Z(t)=Z(-t)$ and $|Z(t)|=|\zeta(\frac{1}{2}+it)|$. Therefore the non-trivial zeros of the Riemann zeta-function in the critical line $\Re\,(s)=\frac{1}{2}$ correspond to real zeros of $Z(t)$.  
For instance, there are other ways to construct modified zeta-functions, as for example, introducing a sharp cut-off in the Euler product. See, for example, reference \cite{leclair}.

In a quite different scenario, in order to discuss analytic regularization procedures to obtain the renormalized vacuum energy of quantum fields in ultrastatic spacetimes with boundary, it was introduced mixed cut-offs in the vacuum energy \cite{cutoff1,cutoff2}.  
Inspired by this idea, here we first modify the Riemann zeta-function introducing a smooth cut-off. Therefore, let us start defining a regularized zeta-function $F(s,\lambda)$ as 
\begin{equation}
F(s,\lambda)=\sum_{n=1}^{\infty}\frac{1}{n^{s}}e^{-\lambda\pi n^{2}},
\end{equation}
where $s=\sigma+i t$; $\sigma$,\,$t$\, and $\lambda\, \in{\mathbb{R}}$, $\lambda>0$. Notice that $F(s,\lambda)$ satisfies the following partial differential equation 
\begin{equation}
\frac{\partial}{\partial\lambda}F(s,\lambda)=-\pi F(s-2,\lambda).
\end{equation}
Discussing possible generalizations to the Riemann zeta-function, Hilbert investigated a similar functional equation \cite{hilbert} . Following the same steps that have been discussed to derive the analytic extension of the Riemann zeta-function, we obtain the following equation for this modified zeta-function
\begin{equation}
\pi^{-\frac{s}{2}}\Gamma\biggl(\frac{s}{2}\biggr)F(s,\lambda)=A(s,\lambda)+B(s,\lambda)+C(s,\lambda)+D(s,\lambda),
\end{equation}
where
\begin{equation}
A(s,\lambda)=\int_{1+\lambda}^{\infty}\,dy\,(y-\lambda)^{\frac{1}{2}(s-2)}\psi(y)
\end{equation}
\begin{equation}
B(s,\lambda)=\int_{\lambda}^{1+\lambda}\,dy\,(y-\lambda)^{\frac{1}{2}(s-2)}\frac{1}{\sqrt{y}}\psi\biggl(\frac{1}{y}\biggr)
\end{equation}
\begin{equation}
C(s,\lambda)=-\frac{1}{2}\int_{\lambda}^{1+\lambda}\,dy\,(y-\lambda)^{\frac{1}{2}(s-2)}
\end{equation}
and finally
\begin{equation}
D(s,\lambda)=\frac{1}{2}\int_{\lambda}^{1+\lambda}\,dy\,(y-\lambda)^{\frac{1}{2}(s-2)}\frac{1}{\sqrt{y}}.
\end{equation}
Instead of discussing regularized zeta-functions with sharp or smooth cut-offs, we propose a modification in the integral representation of the zeta-function, in order to answer the following questions. What modifications are possible in order to obtain a generalized Riemann functional equation in the critical strip? Also, with one regularized zeta-function that satisfies a generalized functional equation, can one show connections between diffusion processes and this regularized zeta-function? 

From the foregoing  discussion, in the analytic extension of the Riemann zeta-function $\zeta(s)$ from $\Re\,(s)>1$ to the whole complex plane,  a particular case of the Jacobi theta-function, the $\Theta(v)$ appears. The main idea of this work is to discuss modifications in one integral representation of the zeta-function introducing functions that exhibit the symmetry $x\mapsto 1/x$. We start defining a modified zeta-function $\zeta_{h}(s,\lambda)$,  for  $\lambda \in{\mathbb{R}}$  and $\lambda >0$, constructed with an arbitrary continuous function $h(x;\lambda)$ such that $h(\frac{1}{x};
\lambda)=h(x;\lambda)$, and $\lim_{\lambda\rightarrow\,0}h(x;\lambda)=1$.  We have
\begin{equation}
\pi^{-\frac{s}{2}}\,\Gamma\biggl(\frac{s}{2}\biggr)\zeta_{h}(s,\lambda)=
\int_{0}^{\infty}dx\, \psi(x)\,h(x;\lambda)\,x^{\frac{s}{2}-1}
\label{zz33}
\end{equation}
Following Riemann's procedure,  we have
\begin{align}
\pi^{-\frac{s}{2}}\,\Gamma\biggl(\frac{s}{2}\biggr)\zeta_{h}(s,\lambda)&=
\frac{1}{2}\int_{1}^{\infty}dx \,h(x;\lambda)\,x^{-\frac{1}{2}(s+1)}-\frac{1}{2}\int_{0}^{1}dx \,h(x;\lambda)\,x^{\frac{1}{2}(s-2)}\nonumber\\
&+\int_{1}^{\infty}dx \,\psi(x)\,h(x;\lambda)\left(x^{\frac{1}{2}(s-2)}+x^{-\frac{1}{2}(s+1)}\right). 
\label{zz44}
\end{align}

We can use the same idea to calculate $\pi^{-\frac{(1-s)}{2}}\Gamma\biggl(\frac{1-s}{2}\biggr)\zeta_{h}(1-s,\lambda)$. Subtracting both expressions we have
\begin{align}\label{eq:funcionallambda}
\pi^{-\frac{(1-s)}{2}}
&\Gamma\biggl(\frac{1-s}{2}\biggr)\zeta_{h}(1-s,\lambda)+\frac{1}{2}\int_{0}^{\infty}dx \, h(x;\lambda)\,x^{-\frac{1}{2}(s+1)}
=\nonumber\\
&\pi^{-\frac{s}{2}}\,\Gamma\biggl(\frac{s}{2}\biggr)\zeta_{h}(s,\lambda)+
\frac{1}{2}\int_{0}^{\infty}dx \,h(x;\lambda)\,x^{\frac{1}{2}(s-2)}.
\end{align}
The above equation is a generalization for the well known functional equation satisfied by the Riemann zeta-function and possesses the same symmetry $s\to (1-s)$ which is present in the Riemann functional equation inside the critical strip. For example we can adopt the functional form $h(x;\lambda)=e^{-\lambda\bigl(f(x)+f(\frac{1}{x})\bigr)}$, for a continuous function $f(x)$. In the next section, using a particular form of the function $f(x)$, a functional equation involving modified Bessel function of second kind is obtained.

\section{A exponential cut-off with the symmetry $x\mapsto 1/x$}\label{Sec:modified2}

Following the symmetry of this theta-function, let us define a regularized zeta-function, i.e., a zeta-function with a cut-off $\zeta(s,\lambda)$,  for  $\lambda \in{\mathbb{R}}$  and $\lambda >0$. For $\lambda=0$,$\,\,\,$ $\Re\, (s)=\sigma >1$. Introducing a cut-off that exhibit the symmetry $x\mapsto 1/x$, we define a regularized zeta-function $\zeta(s,\lambda)$ that can be written as
\begin{equation}
\pi^{-\frac{s}{2}}\,\Gamma\biggl(\frac{s}{2}\biggr)\zeta(s,\lambda)=
\int_{0}^{\infty}dx \,x^{\frac{s}{2}-1}\sum_{n=1}^{\infty}\,e^{-n^{2}\,\pi\,x}e^{-\lambda\bigl(x+\frac{1}{x}\bigr)}. 
\label{zz3}
\end{equation}
Again, following Riemann's procedure, one can show that
\begin{align}
\pi^{-\frac{s}{2}}\,\Gamma\biggl(\frac{s}{2}\biggr)\zeta(s,\lambda)&=
\frac{1}{2}\int_{0}^{1}dx \,e^{-\lambda\left(x+\frac{1}{x}\right)}x^{\frac{1}{2}(s-3)}-\frac{1}{2}\int_{0}^{1}dx \,e^{-\lambda\left(x+\frac{1}{x}\right)}x^{\frac{1}{2}(s-2)}\nonumber\\
&+\int_{1}^{\infty}dx \,\psi(x)e^{-\lambda\left(x+\frac{1}{x}\right)}\left(x^{\frac{1}{2}(s-2)}+x^{-\frac{1}{2}(s+1)}\right). 
\label{zz4}
\end{align}
Our aim is to find one generalization of a well known functional equation satisfied by the Riemann zeta-function. Let us discuss this generalized functional equation. As the cut-off is $x\mapsto 1/x$-invariant, performing the substitution $x\mapsto 1/x$ in the first integral that appears in equation (\ref{zz4}), one finds
\begin{align}
\,\,\pi^{-\frac{s}{2}}\,\Gamma\biggl(\frac{s}{2}\biggr)\zeta(s,\lambda)&=
\frac{1}{2}\int_{1}^{\infty}dx \,e^{-\lambda\left(x+\frac{1}{x}\right)}x^{-\frac{1}{2}(s+1)}-\frac{1}{2}\int_{0}^{1}dx \,e^{-\lambda\left(x+\frac{1}{x}\right)}x^{\frac{1}{2}(s-2)}\nonumber\\
&+\int_{1}^{\infty}dx \,\psi(x)e^{-\lambda\left(x+\frac{1}{x}\right)}\left(x^{\frac{1}{2}(s-2)}+x^{-\frac{1}{2}(s+1)}\right). 
\label{zz5}
\end{align}
We can use the same idea to compute $\pi^{-\frac{(1-s)}{2}}\Gamma\bigl(\frac{1-s}{2}\bigr)\zeta(1-s,\lambda)$. We obtain
\begin{align}\label{zz6}
\pi^{-\frac{(1-s)}{2}}
\Gamma\biggl(\frac{1-s}{2}\biggr)\zeta(1-s,\lambda)&=\frac{1}{2}\int_{1}^{\infty}dx \,e^{-\lambda\left(x+\frac{1}{x}\right)}x^{\frac{1}{2}(s-2)}-\frac{1}{2}\int_{0}^{1}dx \,e^{-\lambda\left(x+\frac{1}{x}\right)}x^{-\frac{1}{2}(s+1)}\nonumber\\
&+\int_{1}^{\infty}dx \,\psi(x)e^{-\lambda\left(x+\frac{1}{x}\right)}\left(x^{-\frac{1}{2}(s+1)}+x^{\frac{1}{2}(s-2)}\right). 
\end{align}
Taking the difference between equation \eqref{zz6} and \eqref{zz5}, we get
\begin{align}\label{eq:funcionallambda}
\pi^{-\frac{(1-s)}{2}}&\Gamma\biggl(\frac{1-s}{2}\biggr)\zeta(1-s,\lambda)+\frac{1}{2}\int_{0}^{\infty}dx \,e^{-\lambda\left(x+\frac{1}{x}\right)}x^{-\frac{1}{2}(s+1)}=\nonumber\\
&\pi^{-\frac{s}{2}}\,\Gamma\biggl(\frac{s}{2}\biggr)\zeta(s,\lambda)+
\frac{1}{2}\int_{0}^{\infty}dx \,e^{-\lambda\left(x+\frac{1}{x}\right)}x^{\frac{1}{2}(s-2)}.
\end{align}
Using the following integral representation of the modified Bessel functions of second kind \cite{AS}
\begin{equation}\label{eq:Kintegral}
\int_{0}^{\infty}\, dx\, x^{\nu-1}e^{-\frac{\beta}{x}-\gamma\,x}=2\left(\frac{\beta}{\gamma}\right)^{\frac{\nu}{2}}K_{\nu}\left(2\sqrt{\beta\gamma}\right)\,\,\,\, (\Re\,(\beta)>0\,\,,\Re\,(\gamma)>0)
\end{equation}
we can rewrite equation \eqref{eq:funcionallambda} as
\begin{equation}\label{eq:functionallambda2}
 \pi^{-\frac{(1-s)}{2}}
\Gamma\biggl(\frac{1-s}{2}\biggr)\zeta(1-s,\lambda)+K_{\frac{1-s}{2}}\left(2\lambda\right)=\pi^{-\frac{s}{2}}\,\Gamma\biggl(\frac{s}{2}\biggr)\zeta(s,\lambda)+K_{\frac{s}{2}}\left(2\lambda\right).
\end{equation} 
The above equation is the first main result of the paper. We obtained a possible generalization of the original functional equation given by equation (\ref{eq:Riemannoriginal}) which possesses the same symmetry $s\to (1-s)$ which is present in the functional equation constructed with the Riemann zeta-function inside the critical strip. For $\lambda>0$ the equation (\ref{eq:functionallambda2}) is defined inside the critical strip. It is not difficult to show that
\begin{align}
\pi^{-\frac{s}{2}}\,\Gamma\biggl(\frac{s}{2}\biggr)&\zeta(s,\lambda)=\frac{1}{2}\int_{0}^{1}dx \,\left(x^{\frac{1}{2}(s-3)}-x^{\frac{1}{2}(s-2)}\right)e^{-\lambda\left(x+\frac{1}{x}\right)}\nonumber \\
&-\int_{0}^{1}dx \,\psi(x)\left(x^{\frac{1}{2}(s-2)}+x^{-\frac{1}{2}(s+1)}\right)e^{-\lambda\left(x+\frac{1}{x}\right)}\nonumber \\
&+2\sum_{n=1}^{\infty}\biggl(\frac{\lambda}{\lambda+n^{2}\pi}\biggr)^{\frac{s}{4}}K_{\frac{s}{2}}\biggl(2\sqrt{\lambda^{2}+\lambda\,n^{2}\pi}\biggr)+2\sum_{n=1}^{\infty}\biggl(\frac{\lambda}{\lambda+n^{2}\pi}\biggr)^{\frac{1-s}{4}}K_{\frac{1-s}{2}}\biggl(2\sqrt{\lambda^{2}+\lambda\,n^{2}\pi}\biggr). 
\label{zz7}
\end{align}
This representation is defined for the whole $s$-complex plane. We would like to point out that these series involving Bessel function were discussed in reference \cite{watson}. 
Let us define the entire function $\xi(s,\lambda)$ by 
\begin{equation}\label{entire}
\xi(z,\lambda)=\frac{1}{2}\pi^{s/2}\Gamma\biggl(\frac{s}{2}\biggr)\zeta(s,\lambda)s(s-1).
\end{equation}
Therefore we have
\begin{equation}\label{entire0}
\xi(s,\lambda)=s(s-1)\sum_{n=1}^{\infty}\biggl(\frac{\lambda}{\lambda+n^{2}\pi}\biggr)^{\frac{s}{4}}K_{\frac{s}{2}}\biggl(2\sqrt{\lambda^{2}+\lambda\,n^{2}\pi}\biggr).
\end{equation}
For small $\lambda$ we can define $\rho=\sqrt{4\pi\lambda}$ and write
\begin{equation}\label{entire}
\lim_{\rho\rightarrow 0^{+}}\xi(s,\rho)=s(s-1)\sum_{n=1}^{\infty}\biggl(\frac{\rho}{2\pi\,n}\biggr)^{\frac{s}{2}}K_{\frac{s}{2}}\bigl(\rho\,n\bigr).
\end{equation}
Let us define the function $\Omega(s,\lambda)$. We have
\begin{equation}\label{flux}
\Omega(s,\lambda)=\frac{1}{2}s(s-1)\biggl(\pi^{-\frac{s}{2}}\,\Gamma\biggl(\frac{s}{2}\biggr)\zeta(s,\lambda)+K_{\frac{s}{2}}\left(2\lambda\right)\biggr).
\end{equation}
Using the fact that 
\begin{equation}
\lim_{x\rightarrow 0} K_{\nu}(x)\approx \frac{2^{\nu-1}\Gamma(\nu)}{x^{\nu}},
\end{equation}
we see that the above equation is singular in the absence of the cut-off. We also have a generalized functional equation $\Omega(s,\lambda)=\Omega(1-s,\lambda)$.
In the next section we will discuss a new functional equation with only one modified Bessel function of second kind.

\section{Exponential $x^{\alpha}$-cut-off}\label{Sec:modified4}

The aim of this section is to discuss a new functional equation with only one Bessel function after introducing a new cut-off. A modified zeta-function is introduced through a new parameter $\alpha$ $(\alpha\in\mathbb{R}\setminus\left\{0\right\})$ in the exponential cut-off where  $\lambda \in{\mathbb{R}}$  and $\lambda >0$. For $\lambda=0$,$\,\,\,$ $\Re\, (s)=\sigma >1$. We have
\begin{equation}\label{eq:zetacutoff}
 \pi^{-\frac{s}{2}}\Gamma\left(\frac{s}{2}\right)\zeta(s,\lambda,\alpha)=\int_{0}^{\infty}dx\,x^{\frac{1}{2}(s-2)}\sum_{n=1}^{\infty}e^{-n^{2} \pi x}e^{-\lambda\left(x^{\alpha}+\frac{1}{x^{\alpha}}\right)}.
\end{equation}
Performing the same steps as before we get 
\begin{align}\label{eq:xis}
&\pi^{-\frac{s}{2}}\Gamma\left(\frac{s}{2}\right)\zeta(s,\lambda,\alpha)=\nonumber\\
 &-\frac{1}{2}\int_{0}^{1}dx\,x^{\frac{1}{2}(s-2)}e^{-\lambda\left(x^{\alpha}+\frac{1}{x^{\alpha}}\right)}+\frac{1}{2}\int_{1}^{\infty}dx\,x^{-\frac{1}{2}(s+1)}e^{-\lambda\left(x^{\alpha}+\frac{1}{x^{\alpha}}\right)}+\nonumber \\
&+\int_{1}^{\infty}dx\,x^{-\frac{1}{2}(s+1)}\psi\left(x\right)e^{-\lambda\left(x^{\alpha}+\frac{1}{x^{\alpha}}\right)}+\int_{1}^{\infty}dx\,x^{\frac{1}{2}(s-2)}\psi(x)e^{-\lambda\left(x^{\alpha}+\frac{1}{x^{\alpha}}\right)}
 \end{align}
and
\begin{align}\label{eq:xi(1-s)}
&\pi^{-\frac{(1-s)}{2}}\Gamma\left(\frac{1-s}{2}\right)\zeta(1-s,\lambda,\alpha)=\nonumber\\
 &=-\frac{1}{2}\int_{0}^{1}dx\,x^{-\frac{1}{2}(s+1)}e^{-\lambda\left(x^{\alpha}+\frac{1}{x^{\alpha}}\right)}+\frac{1}{2}\int_{1}^{\infty}dx\,x^{\frac{1}{2}(s-2)}e^{-\lambda\left(x^{\alpha}+\frac{1}{x^{\alpha}}\right)}+\nonumber \\
&+\int_{1}^{\infty}dx\,x^{\frac{1}{2}(s-2)}\psi\left(x\right)e^{-\lambda\left(x^{\alpha}+\frac{1}{x^{\alpha}}\right)}+\int_{1}^{\infty}dx\,x^{-\frac{1}{2}(s+1)}\psi(x)e^{-\lambda\left(x^{\alpha}+\frac{1}{x^{\alpha}}\right)}.
 \end{align}
The difference is
\begin{align}\label{eq:funcional1}
&\pi^{-\frac{(1-s)}{2}}\Gamma\left(\frac{1-s}{2}\right)\zeta(1-s,\lambda,\alpha)-
\pi^{-\frac{s}{2}}\Gamma\left(\frac{s}{2}\right)\zeta(s,\lambda,\alpha)=\nonumber\\
&-\frac{1}{2}\int_{0}^{\infty}dx\,x^{-\frac{1}{2}(s+1)}e^{-\lambda\left(x^{\alpha}+\frac{1}{x^{\alpha}}\right)}+\frac{1}{2}\int_{0}^{\infty}dx\,x^{\frac{1}{2}(s-2)}e^{-\lambda\left(x^{\alpha}+\frac{1}{x^{\alpha}}\right)}.
\end{align}
To proceed we use the following formula \cite{GR}:
\begin{equation}
    \int_{0}^{\infty}dx\,x^{\nu-1}\exp\left(-\beta x^{p}-\gamma x^{-p}\right)=\frac{2}{p}\left(\frac{\gamma}{\beta}\right)^{\frac{\nu}{2p}}K_{\frac{\nu}{p}}\left(2\sqrt{\beta\gamma}\right)\;\;\left(\Re\,(\beta)>0,\Re\,(\gamma)>0\right).
\end{equation}
Therefore, we obtain this result, valid for $\Re\,(\lambda)>0$:
\begin{equation}\label{eq:functional1}
\pi^{-\frac{(1-s)}{2}}\Gamma\left(\frac{1-s}{2}\right)\zeta(1-s,\lambda,\alpha)-
\pi^{-\frac{s}{2}}\Gamma\left(\frac{s}{2}\right)\zeta(s,\lambda,\alpha)		
=-\frac{1}{\alpha}K_{\frac{1-s}{2\alpha}}\left(2\lambda\right)+\frac{1}{\alpha}K_{\frac{s}{2\alpha}}\left(2\lambda\right).
\end{equation}
The generalized functional equation is then gotten
\begin{equation}\label{eq:functional2}
\pi^{-\frac{(1-s)}{2}}\Gamma\left(\frac{1-s}{2}\right)\zeta(1-s,\lambda,\alpha)
+\frac{1}{\alpha}K_{\frac{1-s}{2\alpha}}\left(2\lambda\right)=
\pi^{-\frac{s}{2}}\Gamma\left(\frac{s}{2}\right)\zeta(s,\lambda,\alpha)		
+\frac{1}{\alpha}K_{\frac{s}{2\alpha}}\left(2\lambda\right).
\end{equation}
Notice that the symmetry $s\to 1-s$ is preserved. Interestingly, since we have a new parameter $\alpha$, we can suitably choose this value to simplify equation \eqref{eq:functional1}. From the formula  \cite{AS}:
\begin{equation}\label{eq:recurrence}
K_{\nu-1}(z)-K_{\nu+1}(z)=\frac{2\nu}{z}K_{\nu}(z),
\end{equation}
and using the property of modified Bessel of second kind: $K_{\nu}(z)=K_{-\nu}(z)$, equation \eqref{eq:funcional1} can be rewritten as
\begin{equation}
\pi^{-\frac{(1-s)}{2}}\Gamma\left(\frac{1-s}{2}\right)\zeta(1-s,\lambda,\alpha)-
\pi^{-\frac{s}{2}}\Gamma\left(\frac{s}{2}\right)\zeta(s,\lambda,\alpha)		
=\frac{1}{\alpha}\left(-K_{\frac{1-s}{2\alpha}}\left(2\lambda\right)+K_{\frac{-s}{2\alpha}}\left(2\lambda\right)\right).
\end{equation}
For $\alpha=1/4$, the above equation can be recast as
\begin{equation}
\pi^{-\frac{(1-s)}{2}}\Gamma\left(\frac{1-s}{2}\right)\zeta(1-s,\lambda,1/4)-
\pi^{-\frac{s}{2}}\Gamma\left(\frac{s}{2}\right)\zeta(s,\lambda,1/4)	
=2\left(-K_{2(1-s)}\left(2\lambda\right)+K_{-2s}\left(2\lambda\right)\right).
\end{equation}
Now we can use equation \eqref{eq:recurrence} to find the following identity:
\begin{equation}
\pi^{-\frac{(1-s)}{2}}\Gamma\left(\frac{1-s}{2}\right)\zeta(1-s,\lambda,1/4)-
\pi^{-\frac{s}{2}}\Gamma\left(\frac{s}{2}\right)\zeta(s,\lambda,1/4)	
=2\left(\frac{1-2s}{\lambda}\right)K_{1-2s}\left(2\lambda\right).
\end{equation}
Therefore we present a new functional equation which involves only one modified Bessel function of second kind. We demonstrated that is possible to modify one integral representation of the zeta-function by introducing functions that obey the symmetry $x\mapsto 1/x$ and still  obtain functional equations with the symmetry $s\mapsto (1-s)$. In the next section we discuss the similarities  between a regularized zeta-function with two parameters and diffusion processes.

\section{Regularized zeta-functions and diffusion processes}\label{Sec:modified3}

The aim of this section is to discuss connections between regularized zeta-functions and diffusion processes. The integral representation of the original regularized zeta-function is closely related to integral representation of the Bessel functions. These functions also appears in diffusion processes. Being more precise, the Laplace transform of the heat kernel for the Euclidean space $\mathbb{R}^{d}$ is written in terms of Bessel functions.  Therefore it is not surprising that one modified zeta-function and  the Laplace transform of the heat kernel are closed connected. Let us define the concentration of particles at the point $\mathbf{x}\in\mathbb{R}^{d}$ at $t$ by $u(t,\mathbf{x})$. The particle concentration at time $t$, when the initial condition was $u_{0}(x)$ is given by
\begin{equation}\label{eq:diffusion}
\frac{\partial}{\partial t}u(t,\mathbf{x})=
\Delta u(t,\mathbf{x}),\quad u(0+,\mathbf{x})=u_{0}(\mathbf{x}),
\end{equation}
where $\Delta$ is the Laplacian in $\mathbb{R}^{d}$ and the diffusion constant is equal to one for simplicity. A general solution is written as
\begin{equation}\label{eq:diffusionsolution}
u(t,\mathbf{x})=\int \; p_{t}(\mathbf{x}-\mathbf{y})u_{0}(\mathbf{y})\;d\mathbf{y}
\end{equation}
where
\begin{equation}
p_{t}(\mathbf{x}-\mathbf{y})=\frac{1}{(4\pi t)^{d/2}}e^{-\frac{|\mathbf{x}-\mathbf{y}|^{2}}{4 t}},
\end{equation}
is the probability of finding the particle at $\mathbf{y}$ and time $t$, if the particle was released from $\mathbf{x}$ at $t=0$. We introduce the operator $P_{t}$ by:
\begin{equation}\label{eq:Pt}
P_{t}f(\mathbf{x})=\int \; p_{t}(\mathbf{x}-\mathbf{y})f(\mathbf{y})\;d\mathbf{y}.
\end{equation}
For $\alpha\,\, \in{\mathbb{C}}$, the resolvent of $P_{t}$ is given by
\begin{eqnarray}\label{eq:resolventPt}
U^{\alpha}f(\mathbf{x})&=&\int e^{-\alpha t} P_{t}f(\mathbf{x})\;dt=\int e^{-\alpha t} p_{t}(\mathbf{x}-\mathbf{y})f(\mathbf{y})\;dt\;d\mathbf{y} \nonumber \\
&=&\int  u_{\alpha}(\mathbf{x}-\mathbf{y})f(\mathbf{y})\;d\mathbf{y}
\end{eqnarray}
where the $u_{\alpha}(\mathbf{x}-\mathbf{y})$ can be written as
\begin{equation}\label{res} 
u_{\alpha}(\mathbf{x}-\mathbf{y})=\frac{2}{(2\pi)^{\frac{d-2}{2}}} \left(\frac{\sqrt{2\alpha}}{|\mathbf{x}-\mathbf{y}|}\right)^{\frac{d-2}{2}}K_{\frac{d-2}{2}}(\sqrt{2\alpha}|\mathbf{x}-\mathbf{y}|),
\end{equation}
where the Bessel function of second kind has purely imaginary argument.
Now, we can discuss some similarities between one regularized zeta-function and the mathematics of diffusion processes. 
To ensure the connection between the one modified zeta-function and the above expression we need to assume that $\lambda\,\, \in{\mathbb{C}}$. Note that, in this case we are in the scenario of two complex variables \cite{c1,c2}. We have $\zeta(s,\lambda)$,  for  $\lambda,s \in{\mathbb{C}}$.  
It is convenient to express the modified zeta-function as 
\begin{equation}\label{zetabesselseries2}
\zeta(s,\lambda)=\frac{2\pi^{\frac{s}{2}}}{(2\lambda)^{\frac{s}{2}}\Gamma(s/2)}\sum_{n=1}^{\infty}\left(\frac{2\lambda}{\sqrt{1+\frac{n^{2}\pi}{\lambda}}}\right)^{\frac{s}{2}}K_{\frac{s}{2}}\left(2\lambda\sqrt{1+\frac{\,n^{2}\pi}{\lambda}}\right).
\end{equation}
Comparing  equation (\ref{res}) and equation (\ref{zetabesselseries2}) one can notice a similarity between the Laplace transform of the heat kernel for the Euclidean space $\mathbb{R}^{d}$ and the modified zeta-function $\zeta(s,\lambda)$. To be more precise, we shall adopt an alternative approach to discuss the connection between regularized zeta-functions and diffusion processes.

To establish a link between the Laplace transform of heat-kernels and regularized zeta-functions, we may consider other modifications in an integral representation of the Riemann zeta-function introducing arbitrary functions that exhibit the symmetry $t\mapsto 1/t$. In the space $\mathbb{C}^{3}=\mathbb{C}\times\mathbb{C}\times\mathbb{C}$, the Cartesian product of three copies of the complex plane, it is possible to define multivariable zeta-functions. We define the regularized zeta-function $\zeta(s,\lambda_{1},\lambda_{2})$ where $s,\lambda_{1},\lambda_{2}\,\,{\in\mathbb{C}}$ where  $\Re\,(\lambda_{1})>0$ and $\Re\,(\lambda_{2})>0$. For $\Re\,(\lambda_{1})=0$ and $\Re\,(\lambda_{2})=0$, $\,$ $\Re\, (s)=\sigma >1$. Using again the $\Theta(t)$ function defined in equation (\ref{theta})
we can write
\begin{equation}
\pi^{-\frac{s}{2}}\,\Gamma\biggl(\frac{s}{2}\biggr)\zeta(s,\lambda_{1},\lambda_{2})=\frac{1}{2}
\int_{0}^{\infty}dt \,t^{\frac{s}{2}-1}\,\bigl(\Theta(t)-1\bigr)h(t;\lambda_{1},\lambda_{2})
\end{equation}
where a generalized cut-off $h(t;\lambda_{1},\lambda_{2})$ is defined as 
\begin{equation}\label{eq:novo}
h(t;\lambda_{1},\lambda_{2})=\frac{1}{2}\biggl[e^{-\bigl(\lambda_{1}t+\frac{\lambda_{2}}{t}\bigr)}+e^{-\bigl(\frac{\lambda_{1}}{t}+
\lambda_{2}t\bigr)}\biggr].
\end{equation}
This generalized cut-off $h(t;\lambda_{1},\lambda_{2})$  exhibits the symmetry $t\mapsto 1/t$. Consequently the Riemann functional equation can be generalized with the same symmetry $s\to (1-s)$ in the critical strip. We have 
\begin{align}\label{eq:funcionallambda1}
\pi^{-\frac{(1-s)}{2}}
&\Gamma\biggl(\frac{1-s}{2}\biggr)\zeta(1-s,\lambda_{1},\lambda_{2})+\frac{1}{2}\int_{0}^{\infty}dt \, h(t;\lambda_{1},\lambda_{2})\,t^{-\frac{1}{2}(s+1)}
=\nonumber\\
&\pi^{-\frac{s}{2}}\,\Gamma\biggl(\frac{s}{2}\biggr)\zeta(s,\lambda_{1},\lambda_{2})+
\frac{1}{2}\int_{0}^{\infty}dt \,h(t;\lambda_{1},\lambda_{2})\,t^{\frac{1}{2}(s-2)}.
\end{align}
We would like to point out that the study of multivariable zeta-functions is not new in the literature, see $e.g.$ \cite{lagarias2003}. 
For Riemannian manifolds possessing enough symmetries, explicit formulas for the heat kernel exist. 
In the next section we discuss the link between one regularized zeta-function and the Laplace transform of these heat kernels.

\section{The Laplace transform of the heat kernel for $\mathbb{R}^{d}$ and  $\mathbb{H}^{d}$ and regularized zeta-functions}\label{Sec:modified33}

The aim of this section is to establish a connection between regularized zeta-function and the Laplace transform for heat kernels in some manifolds. In an arbitrary connected Riemannian manifold it is possible to define the heat kernel, which is a positive fundamental solution to the diffusion equation. Explicit formulas for the heat kernel $p(t,\mathbf{x},\mathbf{y})$  exist for manifolds possessing enough symmetries. A Riemaniann manifold is called a Cartan-Hadamard manifold if is geodesically complete single connected, non-compact with non-positive sectional curvature. For instance, $\mathbb{R}^{d}$ and  $\mathbb{H}^{d}$, a $d$-dimensional hyperbolic space are Cartan-Hadamard manifolds. Let us start discussing the Euclidean space $\mathbb{R}^{d}$ case. 
Starting from the heat kernel for the Euclidean space $\mathbb{R}^{d}$, $p_{d}(t,\mathbf{x},\mathbf{y}) $
we define the Laplace transform of the heat kernel for $\alpha \in \mathbb{C}$, as $p_{d}(\alpha,\mathbf{x},\mathbf{y})$. We can write
\begin{equation}
{\cal{L}}\bigl(p_{d}(t,\mathbf{x},\mathbf{y})\bigr)=p_{d}(\alpha,\mathbf{x},\mathbf{y})=\int_{0}^{\infty}\,\frac{dt}{(4\pi t)^{d/2}}\,e^{-\bigl(\alpha\,t+\frac{|\mathbf{x}-\mathbf{y}|^{2}}{4 t}\bigr)}.
\end{equation}
Using the generalized cut-off $h(t;\lambda_{1},\lambda_{2})$ defined in equation (\ref{eq:novo}), the regularized zeta-function $\zeta(s,\lambda_{1},\lambda_{2})$ can be written as
\begin{align}
\pi^{-\frac{s}{2}}\,\Gamma\left(\frac{s}{2}\right)\zeta(s,\lambda_{1},\lambda_{2})=&-\frac{1}{4}\int_{0}^{\infty}dt\;t^{\frac{s}{2}-1}e^{-\left(\lambda_{1}t+\frac{\lambda_{2}}{t}\right)}-\frac{1}{4}\int_{0}^{\infty}dt\;t^{\frac{s}{2}-1}e^{-\left(\lambda_{2}t+\frac{\lambda_{1}}{t}\right)}\nonumber \\
&+\frac{1}{4}\int_{0}^{\infty}dt\;t^{\frac{s}{2}-1}\Theta(t)e^{-\left(\lambda_{1}t+\frac{\lambda_{2}}{t}\right)}+\frac{1}{4}\int_{0}^{\infty}dt\;t^{\frac{s}{2}-1}\Theta(t)e^{-\left(\lambda_{2}t+\frac{\lambda_{1}}{t}\right)}.
\end{align}
Using equation \eqref{eq:Kintegral}, the last two integrals can be written as a modified Bessel functions of second kind. We find
\begin{align}\label{eq:zetadiffusion}
\pi^{-\frac{s}{2}}\Gamma\left(\frac{s}{2}\right)\zeta(s,\lambda_{1},\lambda_{2})&=-\frac{1}{4}\int_{0}^{\infty}
dt\;t^{\frac{s}{2}-1}e^{-\left(\lambda_{1}t+\frac{\lambda_{2}}{t}\right)}-\frac{1}{4}\int_{0}^{\infty}dt\;t^{\frac{s}{2}-1}e^{-\left(\lambda_{2}t+\frac{\lambda_{1}}{t}\right)}\nonumber \\
&+\frac{1}{2}\sum_{n \in\,\mathbb{Z}}
\left(\frac{\lambda_{2}}{\lambda_{1}+n^{2}\pi}\right)^{\frac{s}{4}}K_{\frac{s}{2}}\left(2\sqrt{\lambda_{2}(\lambda_{1}+n^{2}\pi)}\right)\nonumber \\
& +\frac{1}{2}\sum_{n \in\,\mathbb{Z}}
\left(\frac{\lambda_{1}}{\lambda_{2}+n^{2}\pi}\right)^{\frac{s}{4}}K_{\frac{s}{2}}\left(2\sqrt{\lambda_{1}(\lambda_{2}+n^{2}\pi)}\right).
\end{align}
It is quite difficult to identify diffusion processes for all contributions in the the above representation for the regularized zeta-function.
In this representation of  $\zeta(s,\lambda_{1},\lambda_{2})$ the first integral represents the usual diffusion, i.e., the Laplace transform for the heat kernel for the Euclidean space $\mathbb{R}^{d}$ when we make the following identifications
\begin{equation}
\lambda_{1}=\alpha,\;\;\lambda_{2}=\frac{|\mathbf{x}-\mathbf{y}|^{2}}{4}, \;\;s=2-d,
\end{equation}
for $\alpha$,\,$d$ $\in \mathbb{C}$. This procedure is quite similar to the dimensional regularization, where to regularize divergent integrals in field theory one assume a complex number of dimension \cite{di1,di2,di3,lei}.
To proceed let us show that the regularized zeta-function 
describes diffusion processes. Using the above identifications, the  transformation $t\mapsto 1/t$ and properties for the $\Theta(t)$ we can write 
\begin{align}
\pi^{\frac{d}{2}-1}\,\Gamma\left(1-\frac{d}{2}\right)&\zeta\biggl(2-d,\alpha,\frac{|\mathbf{x}-\mathbf{y}|^{2}}{4}\biggr)=-\frac{1}{4}\int_{0}^{\infty}\;
\frac{dt}{t^{\frac{d}{2}}}\,
e^{-\left(\alpha t+\frac{|\mathbf{x}-\mathbf{y}|^{2}}{4t}\right)}-\frac{1}{4}\int_{0}^{\infty}\;
\frac{dt}{t^{(2-\frac{d}{2})}}\,
e^{-\left(\alpha t+\frac{|\mathbf{x}-\mathbf{y}|^{2}}{4t}\right)}
\nonumber \\
&+\frac{1}{4}
\int_{0}^{\infty}
\frac{dt}{t^{\frac{1}{2}(1+d)}}\sum_{n \in\,\mathbb{Z}}
e^{-\left(\alpha t+\frac{|\mathbf{x}-\mathbf{y}|^{2}+4\pi n^{2}}{4t}\right)}+\frac{1}{4}
\int_{0}^{\infty}\frac{dt}{t^{(2-\frac{d}{2})}}\sum_{n \in\,\mathbb{Z}}
e^{-\left(\alpha t+\frac{|\mathbf{x}-\mathbf{y}|^{2}+4\pi n^{2}}{4t}\right)}.
\end{align}
The above equation is the second main result of the paper. We have shown that if we define a regularized zeta-function  $\zeta\bigl(2-d,\alpha,\frac{|\mathbf{x}-\mathbf{y}|^{2}}{4}\bigr)$ that satisfies a generalized functional equation, one can show that this regularized zeta-function is connected with the Laplace transform of the heat kernels for the Euclidean space $\mathbb{R}^{d}$.  Let us study the above equation in arbitrary spatial dimensions. For the case $\Re\,(d)=1$ and  $\Im\,(d)=0$, the first term in the right hand side of the above equation describes one-dimensional diffusion. The second term in the right hand side of the above equation describes a diffusion in the usual three-dimensional space. The other two terms describes effective diffusion in $d=2$ and $d=3$. Note that for each term of the series we have also that the square of the distance between two points is modified by $4\pi n^{2}$. For the case $\Re\,(d)=2$ and  $\Im\,(d)=0$, three terms describe diffusion in $d=2$ and one describes diffusion in $d=3$. Finally, for the case, $\Re\,(d)=3$ and  $\Im\,(d)=0$, the first term in the right hand side describes three-dimensional diffusion and the second term describes a diffusion in an effective one-dimensional space. The other terms describe an effective diffusion in $d=4$ and $d=2$. Also we have also that for these terms the square of the distance between two points is modified by $4\pi n^{2}$, for each term of the series. 
The regularized zeta-function $\zeta\bigl(2-d,\alpha,\frac{|\mathbf{x}-\mathbf{y}|^{2}}{4}\bigr)$ is expanded in terms of Laplace transform of heat kernels, therefore describing diffusion in different dimensions. We can write
\begin{align}
\pi^{\frac{d}{2}-1}\,\Gamma\left(1-\frac{d}{2}\right)&\zeta\biggl(2-d,\alpha,\frac{|\mathbf{x}-\mathbf{y}|^{2}}{4}\biggr)=-\frac{1}{4}\int_{0}^{\infty}\;
dt\biggl(\frac{1}{t^{\frac{d}{2}}}+\frac{1}{t^{(2-\frac{d}{2})}}\biggr)\,
e^{-\left(\alpha t+\frac{|\mathbf{x}-\mathbf{y}|^{2}}{4t}\right)}
\nonumber \\
&+\frac{1}{4}\sum_{n \in\,\mathbb{Z}}
\int_{0}^{\infty}
dt\biggl(\frac{1}{t^{\frac{1}{2}(1+d)}}+\frac{1}{t^{(2-\frac{d}{2})}}\biggr)
e^{-\left(\alpha t+\frac{|\mathbf{x}-\mathbf{y}|^{2}+4\pi n^{2}}{4t}\right)}.
\end{align}
It is interesting that in the complex plane for $-1 \leq \Re\,(s) \leq 1$ we have $3\geq \Re\,(d) \geq 1$, 
There are some works discussing in field theory a negative dimensional-space  \cite{d1,d2}. Since the regularized zeta-function is defined in all complex-$s$ plane one can go one step forward discussing the region $\Re\,(s)>2$ and its connection with diffusion in a negative dimensional-space.

The same discussion can be made for hyperbolic spaces, since 
in hyperbolic spaces there are exact formulas for the heat kernel. For odd-dimensional space where the geodesic distance between two points is $\rho$, i.e., $\rho=\text{dist}(\mathbf{x},\mathbf{y})$ we get \cite{grigoryan}
\begin{equation}
p_{d}(t,\rho)=\frac{(-1)^{\frac{d-1}{2}}}{(2\pi)^{\frac{d-1}{2}}}\frac{1}{(4\pi\,t)^{\frac{1}{2}}}\biggl(\frac{1}{\sinh\rho}
\frac{\partial}{\partial\,\rho}\biggr)^{\frac{d-1}{2}}e^{-(\frac{d-1}{2})^{2}t-\frac{\rho^{2}}{4t}}.
\end{equation}
In particular for $d=3$ we have
\begin{equation}
p_{3}(t,\rho)=\frac{1}{(4\pi\,t)^{\frac{3}{2}}}\biggl(\frac{\rho}{\sinh\rho}\biggr)e^{-t-\frac{\rho^{2}}{4t}}.
\end{equation}
For even dimensional space there is also a closed expression. A general expression for the heat kernel on the hyperbolic 
space ${\mathbb{H}^{d}_{K}}$ with constant negative curvature $-K^{2}$ can be found in \cite{davies}. Let us define the Laplace transform for the heat kernel in hyperbolic space ${\cal L}\bigl(p_{d}(t,\rho)\bigr)$. We get
\begin{equation}\label{eq:hyperboliclaplace}
{\cal L}\bigl(p_{d}(t,\rho)\bigr)=p_{d}(\alpha,\rho)=\int_{0}^{\infty}dt\, e^{-\alpha\,t}\,p_{d}(t,\rho).
\end{equation}
Since in the hyperbolic space we get the contribution $\frac{\rho}{\sinh\,\rho}$ the connection between one regularized zeta-function and the Laplace transform  for the heat kernel in hyperbolic space is not so clear as in the Euclidean space case. We can write equation (\ref{eq:hyperboliclaplace}) as  
\begin{equation}
\frac{1}{(4\pi)^{-\frac{3}{2}}}\frac{\sinh\rho}{\rho}p_{d}(\alpha,\rho)=\int_{0}^{\infty}dt\,t^{-\frac{3}{2}}e^{-(1+\alpha)t-\frac{\rho^{2}}{4t}}.
\end{equation}
To identify the first term of the modified zeta function (\ref{eq:zetadiffusion}) with a diffusion process in a Cartan-Hadamard space, in $d=3$, we must have 
\begin{equation}
\lambda_{1}=1+\alpha,\;\;\lambda_{2}=\frac{\rho^{2}}{4}, \;\;s=-1.
\end{equation}
Therefore defining modified zeta-functions that lead to generalizations of the Riemann functional equation with the same symmetry $s\to (1-s)$  in the critical strip, we obtained that in the modified zeta-function $\zeta(s,\lambda_{1},\lambda_{2})$ appears contributions that we identify as the Laplace transform 
for the heat kernels in Euclidean and hyperbolic space.

\section{Conclusions and perspectives} \label{Sec:Conclusion and perspectives}
\quad

The modular symmetry is fundamental in order to obtain the analytic extension of the Riemann zeta-function and a functional equation. The main idea of this work is to discuss the consequences of introducing modifications in an integral representation of the Riemann zeta-function using arbitrary functions that exhibit the symmetry $x\mapsto 1/x$.  In this paper we  modified the zeta-function in order to obtain different generalized functions. The first one satisfies a generalized Riemann functional equation. When we define a modified zeta-function $\zeta_{h}(s,\lambda)$,  for  $\lambda \in{\mathbb{R}}$  and $\lambda >0$, constructed with an arbitrary continuous function $h(x;\lambda)$ such that $h(\frac{1}{x};\lambda)=h(x;\lambda)$, and $\lim_{\lambda\rightarrow\,0}h(x;\lambda)=1$, it is possible to find a generalized functional equation. Therefore the existence of a modified functional equation with the symmetry $s\mapsto (1-s)$  is related to the existence of the symmetry  $x\mapsto 1/x$ of arbitrary function introduced to modified the zeta-function. The second one satisfies a functional equation which involves two Bessel functions of second kind. In the study of the non-trivial zeros of the regularized zeta-function we define the function $\Omega(s,\lambda)$ which play the same role as the $\xi(s)$ function in the Riemann scenario with the functional equation $\xi(s)=\xi(1-s)$. We also have a generalized functional equation $\Omega(s,\lambda)=\Omega(1-s,\lambda)$. The third one, defines a functional equation with only one Bessel function of second kind. Also, for the case where $\lambda\,\, \in{\mathbb{C}}$, we present a connection between a regularized zeta-function and the Laplace transform of the kernel of the diffusion equation in a generic  $d$-dimensional Euclidean space.  

A natural continuation of this work is  to discuss other modifications in an integral representation of the Riemann zeta-function introducing arbitrary functions that exhibit the symmetry $x\mapsto 1/x$. This can be done introducing the regularized zeta-function $\zeta(s,\lambda_{1},\lambda_{2},\nu)$ for $\nu\, {\in\mathbb{R}}$ where $\lambda_{1},\lambda_{2}\,\,{\in\mathbb{C}}$, 
where a generalized cut-off $h(t;\lambda_{1},\lambda_{2},\nu)$ is defined as 
\begin{equation}
h(t;\lambda_{1},\lambda_{2})=\frac{1}{2}\biggl[e^{-\bigl(\lambda_{1}t^{\nu}+\frac{\lambda_{2}}{t^{\nu}}\bigr)}+e^{-\bigl(\lambda_{2}t^{\nu}+\frac{\lambda_{1}}{t^{\nu}}\bigr)}\biggr].
\end{equation}
For arbitrary functions that exhibit the symmetry $x\mapsto 1/x$, the Riemann functional equation can be generalized with the same symmetry $s\to (1-s)$. In an arbitrary connected Riemannian manifold it is possible to define the heat kernel, which is a positive fundamental solution to the diffusion equation. Although explicit formulas for the heat kernel exist for a few number of manifolds, in hyperbolic spaces there are exact formulas for the heat kernel. The connection between the Laplace transform of the heat kernel on the hyperbolic space ${\mathbb{H}^{d}_{K}}$ with constant negative curvature $-K^{2}$ \cite{davies} and a regularized zeta-function $\zeta(s,\lambda_{1},\lambda_{2},\nu)$ deserves further investigations. 

It should be noted that for the case with a modified zeta-function with exponential cut-off $\zeta(s,\lambda)$ where  $\lambda \in{\mathbb{R}}$  and $\lambda >0$ we assume that for $\lambda=0$ we must have $\Re\, (s)=\sigma >1$. The crux of the matter is to discuss the limit $\Re\,(\lambda)\rightarrow 0$, for $\Re\, (s)=\sigma \leq 1$, where the polar structure of $I_{1}(s,\lambda)$ and $I_{2}(s,\lambda)$ defined in equation (\ref{zz4}), given respectively by
\begin{equation}
I_{1}(s,\lambda)=\frac{1}{2}\int_{0}^{1}dx \,e^{-\lambda\left(x+\frac{1}{x}\right)}x^{\frac{1}{2}(s-3)}
\end{equation}
and 
\begin{equation}
I_{2}(s,\lambda)=-\frac{1}{2}\int_{0}^{1}dx \,e^{-\lambda\left(x+\frac{1}{x}\right)}x^{\frac{1}{2}(s-2)}
\end{equation}
must be discussed. Next possibility is to discuss the same problem for multivariable zeta-functions.  In the case of the regularized zeta-function $\zeta(s,\lambda_{1},\lambda_{2})$ where $s,\lambda_{1},\lambda_{2}\,\,{\in\mathbb{C}}$ and  $\Re\,(\lambda_{1})>0$ and $\Re\,(\lambda_{2})>0$. For $\Re\,(\lambda_{1})=0$ and $\Re\,(\lambda_{2})=0$, $\,$we assume that $\Re\, (s)=\sigma >1$. The limit $\Re\,(\lambda_{1})\rightarrow 0$ and/or $\Re\,(\lambda_{2})\rightarrow 0$, $\,$for $\Re\, (s)=\sigma \leq 1$, must be further investigated.

\section*{Acknowledgements} 
We would like to thank G. Menezes, A. Hern\'andez and B. F. Svaiter for useful discussions. This work was partially supported by Conselho Nacional de Desenvolvimento Cient\'{\i}fico e Tecnol\'{o}gico - CNPq, 309982/2018-9 (C.A.D.Z.) and 303436/2015-8 (N.F.S.).


\vspace{0.5cm}


\begin{thebibliography}{99}

\bibitem{riem} Riemann B 1859  {\it Monatsberichte d. Preuss. Acad. d. Wissens. Berlin} {\bf 1859:1860} 671 
\bibitem{hardy}  Hardy G H and Wright E M 1956 {\it An introduction to the Theory of Numbers}  (London: Oxford Clarenton Press)
\bibitem{titchmarsh} Titchmarsh E C 1964 {\it The zeta-function of Riemann} (New York: Stechert-Hafner Service Agency)
\bibitem{ingham} Ingham A E 1990 {\it The Distribution of Prime Numbers} (Cambridge: Cambridge University Press)
\bibitem{livro} Davemport H 2000 {\it Multiplicative Number Theory} (Heidelberg: Springer-Verlag)
\bibitem{hejhal} Hejhal D A {\it Journal D'Analyse Mathematique} 1990 {\bf 55} 59
\bibitem{hl} Hardy G H and Littlewood J E 1929 {\it Proc. London Math. Soc.} {\bf 29} 81
\bibitem{nn} Hardy, G H 1914 {\it Proc. London Math. Soc.} {\bf 13}
\bibitem{alexa} Borwein P, Fee G, Ferguson R and van de Waall A 2007 {\it Experimental Mathematics} {\bf 16} 21
\bibitem{gonek} Gonek S M and Ledoan A H 2009 {\it International Mathematics Research Notices}  {\bf 2010} 1775
\bibitem{brown} Heath-Brown D R 2005 Prime Numbers Theory and the Riemann zeta-function {\it Recent Perspectives in Random Matrix Theory and Number Theory} ({\it London Mathematical Society Lecture Notes Series} 322) ed Mezzadri F and Snaith N C (Cambridge: Cambridge University Press)
\bibitem{sierra} Sierra G 2019 {\it Symmetry} {\bf 11} 494 
\bibitem{lagariassuzuki} Lagarias J C and Suzuki M 2006 {\it Journal of Number Theory} {\bf 118} 98
\bibitem{ki} Ki H 2006 {\it Acta Arithmetica} {\bf 124}  197
\bibitem{stn} Julia B 1990 Statistical Theory of Numbers  {\it Number Theory and Physics} 
({\it Springer Proceedings in Physics} vol 47) ed Luck J M, Moussa D and M. Waldschmidt (Berlin: Springer Verlag).
\bibitem{spec1} Spector D 1990  {\it Commun. Math. Phys.} {\bf 127} 239
\bibitem{bakas}  Bakas I and  Bowick M J 1991 {\it J. Math. Phys.} {\bf 32} 1881
\bibitem{spec2}  Spector D 1996 {\it Commun. Math. Phys.} {\bf 177} 13
\bibitem{spec3} Spector D 1998 {\it J. Math. Phys.} {\bf 39}, 1919
\bibitem{prime5} Duenas J G  and  Svaiter N F 2015 \JP A {\bf 48} 315201
\bibitem{mussardo} Mussardo G 1997 The Quantum Mechanical Potential for the Prime Numbers arXiv:cond-mat/9712010
\bibitem{connes} Connes A and Kreimer D 2000 {\it Comm. Math. Phys.} {\bf 210}, 249
\bibitem{rosu}  Rosu H 2003 {\it Mod. Phys. Lett.} {\bf 18} 1205
\bibitem{zyl} Sakhr J,  Bhaduri R K and  van Zyl B 2003 \PR E {\bf 68} 026206
\bibitem{pre}  Schumayer D,  van Zyl B P and  Hutchinson D W 2008 \PR E {\bf 78} 056215
\bibitem{prime1} Menezes G and  Svaiter N F 2012 Quantum Field Theories and Prime Numbers Spectrum arXiv:1211.5198 [math-ph]
\bibitem{prime2} Menezes G,  Svaiter B F and Svaiter N F 2013 {\it Int. Jour. Mod. Phys.} {\bf A28} 1350128
\bibitem{Frontiers} Cartier P, Julia B, Moussa P and Vanhove P 2006 {\it Frontiers in Number Theory, Physics and Geometry I} (Berlin:Springer Berlin)
\bibitem{Daniel} Schumayer D and Hutchinson D A W 2011 \RMP {\bf 83} 307 
\bibitem{prime3}  Duenas J G and Svaiter N F 2014 {\it Int. Jour. Mod. Phys.} {\bf A29} 1450051
\bibitem{prime4}  Duenas J G,  Svaiter N F and Menezes G 2014 {\it Int. Jour. Mod. Phys.} {\bf A29} 1450182 
\bibitem{leclair} LeClair A 2016 Riemann Hypothesis and Random Walks: the Zeta case arXiv:1601.00914 [math.NT]
\bibitem{cutoff1} Svaiter N F and Svaiter B F 1991 {\it J. Math. Phys.} {\bf 32} 175
\bibitem{cutoff2} Svaiter N F and Svaiter B F 1992 \JP A {\bf 25} 979
\bibitem{hilbert} Hilbert D 1902 \emph{Bull. Amer. Math. Soc.} {\bf 8} 437
\bibitem{watson} Watson G N 1931 {\it The Quarterly Journal of Mathematics} {\bf 2} 298
\bibitem{GR} Gradshteyn I S and Ryzhik I M 2000 \emph{Table of Integrals, Series, and Products} (London: Academic Press)
\bibitem{AS} Abramowitz M and Stegun I A 1964 \emph{Handbook of Mathematical Functions} (Mineola: Dover Publications)
\bibitem{c1} Krantz S G 1992 {\it Function Theory of Several Complex Variables} (Providence USA: AMS Chelsea Publishing)
\bibitem{c2} Gunning R C and Rossi H {\it Analytic Functions of Several Complex Variables}  (Providence USA: AMS Chelsea Publishing)
\bibitem{lagarias2003} Lagarias J C and Rains E 2003 {\it Ann. Inst. Fourier, Grenoble} {\bf 53} 1 
\bibitem{di1} Bollini C G and Giambiagi J J, 1972 {\it Nuovo Cimento} {\bf B12} 20
\bibitem{di2} Ashmore J F, 1972 {\it Nuovo Cimento Lett.} {\bf 4} 289
\bibitem{di3} 't Hooft G and Veltman M 1972 {\it Nucl. Phys.} {\bf B44} 189 
\bibitem{lei} Leibrandt G 1975 {\it Rev. Mod. Phys.} {\bf 47} 849
\bibitem{d1} Halliday I G and Ricotta, 1987 {\it Phys. Lett.} {\bf B} 193
\bibitem{d2} Dunne G V and Halliday I G 1987 {\it Phys. Lett} {\bf B} 247.  
\bibitem{grigoryan} Grigor'yan A and Noguchi M 1998  {\it Bull. London Math. Soc.} {\bf 30} 643
\bibitem{davies} Davies E B and Mandouvalos N 1988 {\it Proc. London Roy. Math. Soc. } {\bf 52} 182

\end{thebibliography}
\end{document}